\documentclass[manuscript]{aastex}

\usepackage{amssymb,amsmath}

\newcommand{\bb}{\begin{equation}}
\newcommand{\ee}{\end{equation}}

\shorttitle{Diffusion in pulsar wind nebulae}
\shortauthors{Vorster and Moraal}

\begin{document}

\title{The Effect of Diffusion on the Particle Spectra in Pulsar Wind Nebulae}

\author{M.J. Vorster and H. Moraal}
\affil{Centre for Space Research, School for
  Physical and Chemical Sciences, North-West University, 2520 Potchefstroom, South Africa}
\email{12792322@nwu.ac.za}

\begin{abstract}
A possible way to calculate particle spectra as a function of position in pulsar wind nebulae is to solve a Fokker-Planck transport equation.  This paper presents numerical solutions to the transport equation with the processes of convection, diffusion, adiabatic losses, and synchrotron radiation included.  In the first part of the paper the steady-state version of the transport equation is solved as a function of position and energy.  This is done to distinguish the various effects of the aforementioned processes on the solutions to the transport equation.  The second part of the paper deals with a time-dependent solution to the transport equation, specifically taking into account the effect of a moving outer boundary.  The paper highlights the fact that diffusion can play a significant role in reducing the amount of synchrotron losses, leading to a modification in the expected particle spectra.  These modified spectra can explain the change in the photon index of the synchrotron emission as a function of position.  The solutions presented in this paper are not limited to pulsar wind nebulae, but can be applied to any similar central source system, e.g. globular clusters.  

\end{abstract}


\section{Introduction}

Relativistic particles transported through interstellar space are subjected to a number of processes, leading to a spatial and temporal variation in the particle energy spectrum.  A natural way to describe the evolution of the energy spectrum, with the aforementioned processes taken into account, is to use a Fokker-Planck transport equation.  Notable examples include the transport equation used to describe the propagation of cosmic rays in the Galaxy \citep{Ginzburg1964}, and the Parker transport equation, used to describe the modulation of cosmic rays in the heliosphere \citep{Parker1965}.

This paper investigates the evolution of a non-thermal particle energy spectrum originating form a central source.  The particles are transported away from the source through convection and diffusion, whilst simultaneously losing energy due to synchrotron radiation, inverse Compton scattering, and adiabatic cooling.  The prototypical example of such a system is the class of objects known as pulsar wind nebulae, with the Crab Nebula being the best-known source.

A supersonic wind originating in the vicinity of the pulsar transports particles and magnetic fields into the surrounding medium.  Although the characteristics of the wind are not fully understood, it is generally believed to consist of highly relativistic leptons (electrons and positrons) \citep[e.g.][and references therein]{Kirk2009}.  When the ram pressure of the pulsar wind reaches an equilibrium with the pressure from the surrounding medium, a termination shock is formed \citep{Rees1974}, capable of accelerating the particles \citep{Kennel1984}.  Downstream of the termination shock the leptons in the pulsar wind interact with the frozen-in magnetic field, leading to the emission of synchrotron radiation ranging from radio to X-rays.  Additionally, the leptons can also inverse Compton scatter photons from the Cosmic Microwave Background Radiation (CMBR) to gamma-ray energies.  This non-thermal emission creates a luminous nebula around the pulsar, referred to as a pulsar wind nebula (PWN).  An important characteristic of PWN'e is that the evolutionary time-scale of the system is often comparable to the age of the system, making it necessary to take dynamic effects, such as the expansion of the system, into account.     

A second possible source type is globular star clusters.  It is believed that non-thermal particles are injected into the cluster by millisecond pulsars located at the centre of the cluster.  The particles propagate away from the centre, producing synchrotron and inverse Compton emission.  The large size of the cluster compared to the compact core of pulsars makes it possible to approximate the central pulsars as a single source \citep[e.g.][]{Venter2009}.

A further possible application of the model is to the bubbles observed below and above the Galactic plane by \emph{WMAP} \citep{Dobler2008}, \emph{Fermi} \citep{Su2010}, and more recently by \emph{Planck} \citep{Planck2012}.  Various explanations for the observed emission have been proposed, ranging from dark matter annihilation to emission from cosmic-ray electrons.  If the latter are the cause of the emission, then the present model can be used to explain some of the observed characteristics of the bubbles.  

Comprehensive analytical solutions of the transport equation with convection, diffusion, and energy changes have been presented by \cite{Lerche1981} for time-dependent point and diffuse sources.  The solutions were obtained for an axis-symmetric, cylindrical coordinate system, with the limitation that the convection velocity must either be constant, or increase linearly from the source.  A partial numerical counterpart to that work is presented in this paper, with the difference that the present model is solved in spherical coordinates without any limitation on the profile of the convection velocity.  The present paper focusses on two main aspects.  It has been observed that the photon index of the X-ray synchrotron emission softens as a function of distance from the pulsar wind termination shock \citep[e.g.][]{Mangano2005, Schock2010}.  The first aim of this paper is to demonstrate that the softening of the spectra can be explained if diffusion is present in PWN'e.  The second aim of the paper is to investigate the effect of a moving outer boundary on the evolution of the particle spectra.

The outline of the paper is as follows: in Section 2 the appropriate time-dependent transport equation is introduced.  In order to reduce the complexity of the problem, spherical symmetry is imposed on the system.  To investigate the effect of the various processes, the transport equation is first solved time-independently, with the results presented in Section 3.  Time-dependent solutions to the transport equation are presented in Section 4, where the effect of a moving outer boundary, as well as a time-dependence in the magnetic field and diffusion coefficient, is taken into account.  Section 5 summarises the main results and discusses planned extensions to the present model.

\section{The Model}

\subsection{The Transport Equation}

The general transport equation for a massive particle species, expressed as a function of momentum $p$ and the omnidirectional distribution function $f(\mathbf{r},p,t)$, is given by \citep[e.g.][]{Ginzburg1964, Parker1965}
\begin{equation}\label{eq:gte}
  \frac{\partial f}{\partial t} + \nabla \cdot \mathbf{S} - \frac{1}{p^2}\frac{\partial}{\partial p}\left(p^2 \left[ \langle\dot{p}\rangle f + D_p \frac{\partial f}{\partial p}\right]\right ) = Q(\mathbf{r},p,t),
\end{equation}
where
\begin{equation}\label{eq:S}
\mathbf{S} = 4\pi p^2\left(C\mathbf{V}f-\mathbf{\underline{K}}\cdot\nabla f\right)
\end{equation}
is the differential current density describing convection and diffusion, with $\mathbf{V}$ denoting the convection velocity and $\mathbf{\underline{K}}$ the diffusion tensor.  The relationship between the omnidirectional distribution function and the more frequently used particle density is given by $N(\mathbf{r},p,t) = 4\pi p^2 f(\mathbf{r},p,t)$, where $N$ is defined as the number of particles per unit volume in the momentum interval $p+dp$.  The expressions for synchrotron and inverse Compton emission spectra are usually stated in terms of $N(\mathbf{r},p,t)$ \citep[e.g.][]{Blumenthal1970, Longair1994}, whereas it is often more convenient to express transport equations in terms of $f(\mathbf{r},p,t)$.

The Compton-Getting coefficient \citep[e.g.][]{Gleeson1968, Gleeson1973}
\begin{equation}
C= -\frac{1}{3} \frac{\partial\ln f}{\partial\ln p}
\end{equation}
takes into account that an observer moving relative to the rest frame in which $f$ is homogeneous and isotropic will measure a modified current density.  This modification produces a spectral effect similar in nature to the Doppler effect on photons.  For a power-law spectrum $f \propto p^{-(\alpha+2)}$, the coefficient has the value $C = (\alpha+2)/3$.

In a spherically expanding medium, energy or momentum losses include the adiabatic loss rate 
\bb\label{eq:ad_stat}
\frac{\langle\dot{p}\rangle_{\rm{ad}}}{p} = -\frac{1}{3}\nabla\cdot\mathbf{V}. 
\ee
However, \cite{Gleeson1978} showed that when the particles reside in a wind with velocity $\mathbf{V}$, this is the momentum loss rate in the frame moving with the wind, while the loss rate in the stationary frame of reference is
\begin{equation}\label{eq:pdot}
\frac{\langle\dot{p}\rangle_{\rm{ad}}}{p} = \frac{1}{3}\mathbf{V}\cdot\frac{\nabla f}{f}.
\end{equation}
The modification of the convective streaming from $\mathbf{V}f$ to $C \mathbf{V}f$, and the adiabatic cooling rate from (\ref{eq:ad_stat}) to (\ref{eq:pdot}), produces effects in the transport equation that cancel each other. \cite{Gleeson1978} showed that when $C$ is dropped in $\mathbf{S}$, the momentum loss rate in (\ref{eq:gte}) transforms from (\ref{eq:pdot}) back to its more familiar form.  It should be noted at this point that the process of first-order Fermi acceleration in shocks is formally included in the transport equation if it is treated as a discontinuous jump in the flow velocity.

Synchrotron radiation and inverse Compton scattering in the Thomson regime have the same momentum dependence, making it possible to describe both non-thermal processes by the single expression
\begin{equation}\label{eq:Thom}
\frac{\langle\dot{p}\rangle_{\rm{n-t}}}{p} = z(\mathbf{r},t)p,
\end{equation}
where
\bb\label{eq:a_coef}
z(\mathbf{r},t) = \frac{4\sigma_{\rm{T}}}{3\left(m_0c\right)^2}\left(U_B+U_{\rm{IC}} \right)
\ee
is a function of the magnetic ($U_B$) and photon ($U_{\rm{IC}}$) energy field densities.  The other constants in (\ref{eq:a_coef}) are $\sigma_{\rm{T}}$, the Thomson scattering cross-section, $m_0$, the rest mass of the particle, and $c$, the speed of light.  In systems where $U_{\rm{IC}} > U_B$, (\ref{eq:Thom}) must be modified to take into account Klein-Nishina effects.  Neglecting convection and diffusion, \cite{Moderski2005} solved (\ref{eq:gte}) time-independently, finding that Klein-Nishina effects are still negligible if $U_{\rm{IC}}/U_B \lesssim 3$.  For the CMBR with an energy density of $0.3$ $\rm{eV}/\rm{cm}^{3}$, this condition is satisfied for an average magnetic field of $\bar{B} > 2$ $\mu\rm{G}$.  

The term $D_p \partial f/\partial p$ describes stochastic or second-order Fermi acceleration in magnetohydrodynamic turbulence, with $D_p$ acting as the momentum diffusion coefficient.  Lastly, any sources or sinks are included in the transport equation through $Q(\mathbf{r},p,t)$.

Inserting the above expressions for $\mathbf{S}$ and $\langle\dot{p}\rangle_{\rm{total}} = \langle\dot{p}\rangle_{\rm{ad}}+\langle\dot{p}\rangle_{\rm{n-t}}$ into (\ref{eq:gte}), imposing spherical symmetry ($\partial f/\partial \phi = \partial f/\partial \theta = 0$) on the system, and neglecting momentum diffusion, $D_p \partial f/\partial p=0$, leads to
\begin{equation}\label{eq:te}
\frac{\partial f}{\partial t} = \kappa\frac{\partial^2 f}{\partial r^2} + \left[\frac{1}{r^2}\frac{\partial}{\partial r}\left(r^2\kappa\right)-V\right]\frac{\partial f}{\partial r} + \left[\frac{1}{3r^2}\frac{\partial}{\partial r}\left(r^2V\right)+z p\right]\frac{\partial f}{\partial \ln p}+4zpf + Q,
\end{equation}
expressed in the spherical coordinate $r$.  In this case $\mathbf{\underline{K}}$ reduces to a single effective isotropic diffusion coefficient $\kappa(p,r,t)$.

\subsection{Scaling of the model}\label{sec:scale}

To keep the model as general as possible, it is useful to scale the variables present in (\ref{eq:te}) to dimensionless ones.  For radial distance, the scaling
\bb
r =\frac{\tilde{r}}{\tilde{r}_{S}}
\ee
is used, where $\tilde{r}$ is the unscaled variable and $\tilde{r}_{S}$ a characteristic length, chosen as the size of the system.  This leads to a system with a scaled radial dimension of $0 < r \le 1$.  A similar scaling is used for momentum
\bb\label{eq:p_scale}
p = \frac{\tilde{p}}{\tilde{p}_{S}},
\ee
with the difference that $\tilde{p}_{S}$ is not chosen as the maximum particle momentum, but rather an intermediate value.  For the present paper, the scaled momentum range is chosen as $10^{-3} \le p \le 10^4$.  It is also possible to scale the particle energy in a similar fashion, $E=\tilde{E}/\tilde{E}_S$.  Since $\tilde{E}=\tilde{p}c$ for relativistic particles, it follows that $E=p$.  For the numerical calculations the scale energy is chosen as $\tilde{E}_S=1$ TeV, with the scaled momentum range translating to the energy range $1 \mbox{ GeV} \le \tilde{E} \le 10$ PeV.  This energy range includes the electron energy range needed to produce the non-thermal emission observed from PWN'e.  

The convection velocity in the system is expressed  as a fraction of the speed of light
\bb
V=\frac{\tilde{V}}{c},
\ee
while the diffusion diffusion coefficient is scaled as
\bb\label{eq:kappa_scale}
\kappa=\frac{\tilde{\kappa}}{\tilde{r}_{S}c}.
\ee
It is useful to define a scaling factor for time, chosen as the light transition time through the system
\bb
t = \frac{c\tilde{t}}{\tilde{r}_{S}}.
\ee
Consequently, the non-thermal energy loss coefficient (\ref{eq:a_coef}) is scaled as
\bb\label{eq:z_scale}
z = \frac{\tilde{z}\tilde{r}_{S}}{c}.
\ee

\subsection{Time-scales}

The amount of energy lost by a particle is directly related to the residence time in the system.  In the absence of diffusion, the convection time-scale is given by
\begin{equation}\label{eq:con_time}
\tau_{\rm{con}} = \int_{r_0}^{r}\frac{dr}{V(r)}.
\end{equation}
Conversely, the diffusion time-scale, as derived by \cite{Parker1965}, is 
\begin{equation}\label{eq:dif_time}
\tau_{\rm{dif}} = \frac{r^2}{6\kappa}.
\end{equation}
It should be noted that diffusion is a stochastic process, and that $\tau_{\rm{dif}}$ represents an average propagation time.   Furthermore, this diffusion time-scale is derived under the assumption that $\kappa$ is independent of $r$, and that the particles undergo negligible energy losses during propagation.  The synchrotron cooling time is estimated to be \citep[e.g.][]{Dejager2009}
\bb\label{eq:tau_syn}
\tilde{\tau}_{\rm{syn}} \sim \frac{8.4\times 10^3}{B_{\mu\rm{G}}^2E_{\rm{TeV}}} \mbox{ kyr}.
\ee

\subsection{Parameter values in the model}\label{sec:parameter}

Magnetohydrodynamic (MHD) simulations show a complicated flow and magnetic field structure in PWN'e \citep[e.g.][]{Komissarov2003, Delzanna2004}.  However, in the present one-dimensional problem a radial flow and  azimuthal magnetic field is assumed, similar to the models of (e.g.) \cite{Rees1974} and \cite{Kennel1984}.  The latter assumption follows from the result that a radial plasma wind originating from a magnetised, rotating object leads to an Archimedean spiral structure in the magnetic field \citep[e.g.][]{Parker1965}.  As a result of the large rotational velocity of pulsars, the spiral structure can be approximated to a high degree as purely azimuthal \citep[e.g.][and references therein]{Kirk2009}.

It is further assumed that the ideal MHD limit
\begin{equation}\label{eq:MHD_limit}
\nabla \times \mathbf{V} \times \mathbf{B} = 0
\end{equation}
holds, allowing one to relate the radial profiles of the velocity and magnetic field.  With the chosen flow and magnetic field structure, (\ref{eq:MHD_limit}) reduces to
\begin{equation}\label{eq:MHD_limit_reduce}
VBr=\mbox{ constant}=V_0B_0r_0.
\end{equation}

The convective flow downstream of the pulsar wind termination shock is subsonic, with the sound speed in a magnetised plasma being of the order $c/\sqrt{3}$ \citep[e.g.][]{Reynolds1984}.  The inner boundary of the computational domain is placed at the termination shock, where the velocity is chosen as $\tilde{V}_0=0.3c$.  Steady-state MHD simulations \citep{Kennel1984} find that the radial velocity profile depends on the ratio of electromagnetic to particle energy, $\sigma$.  When $\sigma=1$, the velocity remains almost independent of $r$.  When $\sigma=0.01$, the velocity falls off as $V \propto 1/r^2$ close to the termination shock, but decelerates at one termination shock radius, and approaches a constant value at a radius that is a hundred times larger than the termination shock radius.  For the current model a $V \propto 1/r^{0.5}$ profile is chosen.  The same profile has been derived by \cite{Vanetten2011} for HESS J1825-137. 

The chosen velocity profile, together with (\ref{eq:MHD_limit_reduce}), implies the magnetic profile $B \propto 1/r^{0.5}$, similar to the scaling $B\propto 1/r^{0.7}$ derived by \cite{Vanetten2011}.  Although the radial dependencies of $V$ and $B$ may be more complicated, the chosen profiles lead to a decrease in the velocity and magnetic field qualitatively similar to the large-scale structure calculated by \cite{Kennel1984}.  

For the magnetic field it is difficult to specify a characteristic value for PWN'e.  The Crab Nebula has an average magnetic field strength of $\bar{B}\sim 300$ $\mu$G \citep[e.g.][]{Trimble1982}, whereas it is estimated that the Vela PWN has a lower value of $\bar{B}\sim 3$ $\mu$G \citep[e.g.][]{Dejager2009}.  For the numerical simulations, the value $\tilde{B}_0=350$ $\mu$G is chosen as the magnetic field strength at the inner boundary.  This is again similar to the value $\tilde{B}_0\approx 400$ $\mu$G derived for HESS J1825-137 \citep{Vanetten2011}.    

It is generally assumed \citep[e.g.][]{Lerche1981} that the radial ($\kappa_r$) and momentum ($\kappa_p$) dependence of the diffusion coefficient can be separated, $\kappa(r,p)=\kappa_0 \kappa_r\kappa_p$, with $\kappa_0$ a constant.  Diffusion occurs as a result of particle interaction with irregularities in the magnetic field, and experience with the propagation of cosmic rays in the heliosphere shows that the radial dependence of the diffusion coefficient can be modelled as $\kappa_r \propto 1/B$ \citep[e.g.][and references therein]{Caballero2004}.  This leads to the radial dependence $\kappa_r \propto r^{0.5}$ in the model.  It has further been found that the momentum dependence of particles in the heliosphere scales as $\kappa_p \propto p$ \citep[e.g.][and references therein]{Caballero2004}.  The momentum dependence $\kappa_p = \tilde{p}/\tilde{p}_S$ is thus chosen in the model.  

To account for the loss of particles with $\tilde{E} > 100$ GeV from the gamma-ray emission region in the Vela PWN, a spatially-independent diffusion coefficient $\tilde{\kappa}_0 \sim 10^{26}$ $\rm{cm}^2/s$ has been estimated by \cite{Hinton2011}.  A similar $\tilde{\kappa}_0$ has also been derived by \cite{Vanetten2011} to explain TeV gamma-ray emission from the source HESS J1825-137.  For a PWN size $\tilde{r}_{S}=10$ pc and $\tilde{E}_{S}=\tilde{p}_{S}c=1$ TeV, the coefficient $\kappa$ has the same value as the observationally derived value when $\kappa_0=10^{-3}$ and $\tilde{E}=100$ GeV.  For the purpose of illustration, the smaller value $\kappa_0=4\times 10^{-5}$ is chosen at the inner boundary, and is motivated by the scaling $\kappa \propto 1/B$.  Since the average magnetic field in the model is larger than $\bar{B}$ in the Vela PWN, a smaller diffusion coefficient is appropriate.  With the radial scaling taken into account (fixed by the radial profile of the magnetic field), the diffusion coefficient varies from $\kappa =4\times 10^{-5}E$ at the inner boundary, to $\kappa=4 \times 10^{-4}E$ at the outer boundary.

Lastly, the particle energy spectrum injected by the source is chosen as $f(p,r_0)\propto p^{-4}$, or in terms of the differential particle density $N\propto p^{-2}$.  This is the index of the particle spectrum needed to explain the X-ray emission observed in the vicinity of the termination shock \citep[e.g.][and references therein]{Gaensler2006}.  Since the particles are relativistic ($\tilde{E}=\tilde{p}c$), and with the scaling of the parameters described in Section \ref{sec:scale}, the source spectrum can also be expressed as $N\propto \tilde{E}^{-2}$.

\section{Steady-state Solutions to the Transport Equation}

The steady-state version of (\ref{eq:te}), i.e. $\partial f/\partial t=0$, is solved using the second-order accurate \emph{Crank-Nicolson} numerical scheme.  The model used in this section is a derivative of the one developed by \cite{Steenkamp1995}, and used by \cite{Steenberg1996} to solve the Parker transport equation describing the modulation of cosmic rays in the heliosphere.

In this steady-state version, momentum plays the numerical role usually assigned to time, and is thus used as the stepping parameter of the parabolic equation.  Because particles migrate downwards in energy or momentum space as a result of energy losses, the numerical scheme steps from high to low momentum values in logarithmic steps.  Particles propagating away from the source immediately lose energy, and as a result particles with momentum $p_{\rm{max}}$ can only exist at the source located at $r_0=0.01$, leading to the "initial" condition $f(p_{\rm{max}},r)=0$ for all $r > r_0$.  For the solutions presented in this section, the radial grid is divided into a 1000 steps between $r_0 \le r\le 1$, with $\Delta r=0.001$, while the stepping parameter ranges between $10^{-3} \le p \le 10^4$, with step size $\Delta \ln p = -0.02$.

The number of particles per momentum interval that flows through the inner boundary must be equal to the total number of particles produced per time and momentum interval, $Q=Q^*p^{-(\alpha+2)}$, with $Q^*$ a normalisation constants.  This implies that
\bb\label{eq:bound_flux}
\oint\mathbf{S}\cdot d\mathbf{A} = Q,
\ee
where $d\mathbf{A}=r_0^2\sin\theta d\theta d\phi \mathbf{e}_r$ is the surface element, and $\mathbf{S}$ is given by (\ref{eq:S}).  Integrating (\ref{eq:bound_flux}) leads to the inner boundary condition
\begin{equation}\label{eq:bound}
CV_0f - \kappa_0\frac{\partial f}{\partial r} = Q^*\frac{1}{4\pi r_0^2}p^{-(\alpha+2)}
\end{equation}    
that is solved simultaneously with (\ref{eq:te}).  A similar boundary condition was derived by \cite{Ng1975} to describe the evolution of the energy spectrum of cosmic rays produced by solar flares.  In the absence of diffusion, (\ref{eq:bound}) reduces to
\bb\label{eq:bound_no_kappa}
f(r_0,p_0) = \frac{Q^*}{CV_0}\frac{1}{4\pi r_0^2}p^{-(\alpha+2)}.
\ee  
To simulate particles escaping from the system, the free-escape (\emph{Dirichlet}) condition $f(r=1,p)=0$ is imposed at the outer boundary.


\subsection{Convection-only and convection-synchrotron}

\begin{figure*}[!t]
   \centering
  \includegraphics[height=10cm]{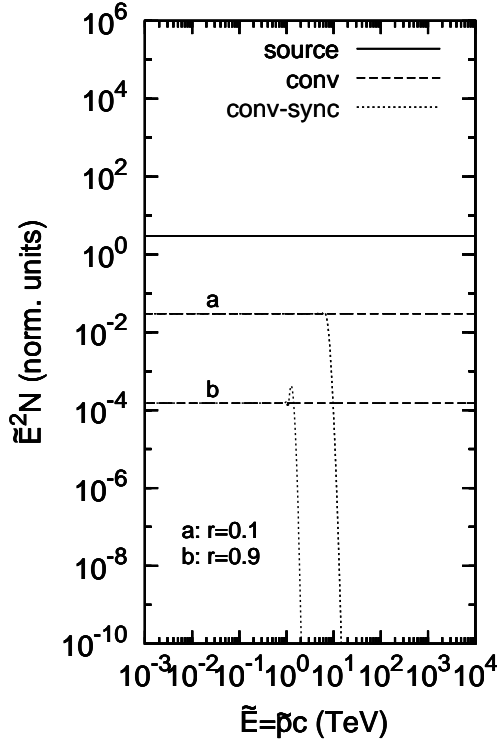}
   \caption{Numerical solutions to the one-dimensional, steady-state version of the transport equation, (\ref{eq:te}), for the convection (conv) and convection-synchrotron (conv-sync) scenarios.  The term "convection" is used as short-hand for the phrase "convection with adiabatic losses".  Note that the differential number density $N=4 \pi p^2f$ is plotted, instead of the distribution function $f$.  To highlight characteristics of the spectrum, $N$ is multiplied by $\tilde{E}^2$.  }
              \label{fig:conv-sync}
    \end{figure*}

This section presents two sets of steady-state solutions to (\ref{eq:te}).  The first set includes only the processes of convection and adiabatic losses, while the second set includes convection, together with adiabatic and synchrotron losses.  For convenience, the former scenario will be referred to as the convection scenario, while the latter will be referred to as the convection-synchrotron scenario.  Figure \ref{fig:conv-sync} shows the spectra for the convection scenario, together with the convection-synchrotron scenario.  When diffusion and synchrotron losses are absent, the spectra shift to lower energies as a result of adiabatic losses, while retaining the spectral shape of the source spectrum.  This can be demonstrated analytically as follows: for the steady-state convection scenario with a radially independent $V=V_0$, the transport equation (\ref{eq:te}) simplifies to
\begin{equation}\label{eq:Ana}
\frac{\partial f}{\partial r} - \frac{2}{3r}\frac{\partial f}{\partial \ln p} = 0.
\end{equation}
Using the method of characteristics to solve the above equation, one finds that the solutions to (\ref{eq:Ana}) are constant along the contours
\bb\label{eq:rp_plane}
p_0=p\left(\frac{r}{r_0}\right)^{2/3}
\ee
in the $(r,p)$-plane, valid for any choice of source function.  With the boundary condition (\ref{eq:bound_no_kappa}) taken into account, this leads to the solution \begin{equation}\label{eq:con_ana}
f(r,p) = \frac{Q^*}{4\pi r_0^2V_0}\left[p\left(\frac{r}{r_0}\right)^{2/3}\right]^{-(\alpha+2)}\mathcal{H}\left(r-r_0\right),
\end{equation}
where $\mathcal{H}$ is the Heaviside function.  

The analytical solution (\ref{eq:con_ana}) shows that the spectra at any position in the system have the same $p$-dependence as the source spectrum, with the only difference being that the intensities are reduced.  This behaviour can be ascribed to the fact that the fractional momentum change as a result of adiabatic losses, (\ref{eq:pdot}), is independent of momentum.  The same result has also been found by (e.g.) \cite{Jokipii1979} and \cite{Lerche1981}.  

The analytical result (\ref{eq:con_ana}) is a useful baseline for obtaining an estimate of the accuracy of the numerical scheme.  Using a radially-independent velocity, it was found that the numerical solutions are within $10\%$ of the analytical solutions in the inner part of the system.  The accuracy decreases as a function of position, so that at $r=0.9$ the intensity of the numerical spectrum is a factor two below the analytical result.  When diffusion is absent, the character of (\ref{eq:te}) changes from parabolic to hyperbolic, and the \emph{Crank-Nicolson} method is no longer the most suitable scheme for solving (\ref{eq:te}).  However, at $r=0.9$ the intensity has dropped a factor of $10^5$ compared to the source spectrum, (see Figure \ref{fig:conv-sync}), and a factor of two becomes negligible.  

\begin{figure*}[!t]
   \centering
  \includegraphics[height=10cm]{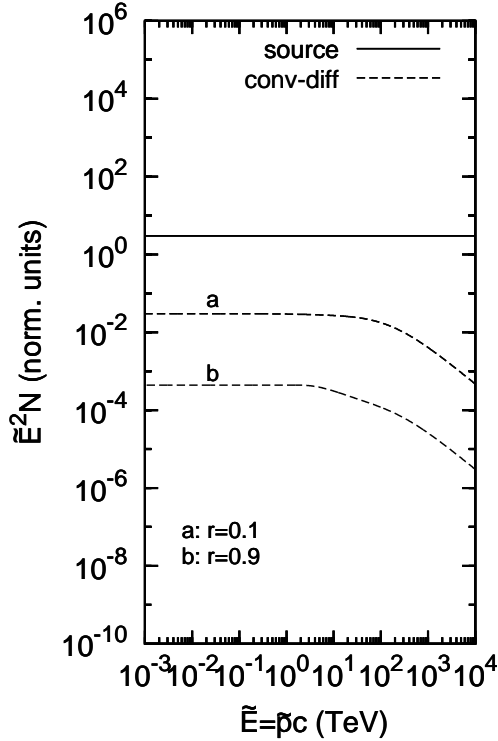}
   \caption{Numerical solutions in the inner and outer part of the system for the convection-diffusion scenario.}
              \label{fig:conv-diff}
    \end{figure*}

Including the effect of synchrotron losses leads to the familiar cut-off that shifts to lower energies with increasing $r$, as shown in Figure \ref{fig:conv-sync}.   Additionally, a small peak is visible just before the cut-off in the spectrum at $r=0.9$.  The most intuitive explanation for this peak would be that synchrotron losses lead to a pile-up of particles at the edge of the cut-off.  Such an effect is indeed expected for the source spectrum $N \propto p^{-\alpha}$ when $\alpha < 2$ \citep[e.g.][]{Longair1994}.  However, the peak in Figure \ref{fig:conv-sync} (where the source spectrum has $\alpha=2$) is not a physical effect, but rather a numerical artefact resulting from the absence of diffusion.  The difference between the intensities of the peak and the adiabatic part of the spectrum ($\tilde{E} \lesssim 1.4$ TeV) is the factor two mentioned in the previous paragraph.  When diffusion is included in the problem, (cf. Figure \ref{fig:conv-diff}), the correct intensity is calculated.  

The energy, $\tilde{E}_{\rm{syn}}$, where the synchrotron break should appear can be estimated by equating the right-hand sides of (\ref{eq:con_time}) and (\ref{eq:tau_syn}), and then solving for $\tilde{E}$.  The value for the magnetic field used in the estimate is the average value in the region ranging from $r_0$ to $r$.  This leads to the estimates $\tilde{E}_{\rm{syn}} \sim 12.6$ TeV at $r=0.1$ and $\tilde{E}_{\rm{syn}} \sim 2.9$ TeV at $r=0.9$.  The break energies in Figure \ref{fig:conv-sync} are located at the lower energies $\tilde{E} \sim 6$ TeV and $\tilde{E} \sim 1.4$ TeV for $r=0.01$ and $r=0.9$, respectively.  A comparison of the numerical results with the order-of-magnitude estimates shows a factor $2$ difference in the break energies, confirming the validity of the numerical scheme.

\subsection{Convection-diffusion}\label{sec:con_dif}

To determine the effect of diffusion on the evolution of the particle spectrum, this section considers steady-state solutions to (\ref{eq:te}), i.e. $\partial f/\partial t=0$, with the processes of convection, diffusion and adiabatic losses included.  The values for the parameters are as discussed in Section \ref{sec:parameter}.  The evolution of the spectra in a convection-diffusion system can be divided into two regimes, as shown in Figure \ref{fig:conv-diff}.  At lower energies convection and adiabatic losses dominate and the spectra retain the shape of the source spectrum.  At higher energies diffusion is more important, leading to an evolution in the spectral index.  With the free-escape (\emph{Dirichlet}) condition imposed at the outer boundary, the diffusion-dominated part of the spectra become softer by one power-law index, $N \propto p^{-3}$, or expressed in terms of energy, $N \propto \tilde{E}^{-3}$ (see Section \ref{sec:scale}).  

This softening of the spectra is typical of escape losses \citep[e.g.][]{Atoyan1995}, and is related to the momentum dependence of the diffusion coefficient.  For the general case $\kappa \propto p^{\lambda}$, the diffusion-dominated part of the spectra is described by $N \propto p^{-\alpha-\lambda}$ \citep[e.g.][]{Jokipii1979,Lerche1981, Atoyan1995}, with $\alpha=2$ and $\lambda=1$ in the present model.  

In the outer part of the system, $r=0.9$, the transition between the two regimes is more gradual, compared to the transition at $r=0.1$.  Particles with $\tilde{E} \lesssim 20$ TeV are transported predominantly by convection in the inner part of the system.  However, since the diffusion coefficient increases with radial distance, the energy where particles are transported by convection has reduced to $\tilde{E} \lesssim 4$ TeV at $r=0.9$.  Particles in the energy range $4 \mbox{ TeV} \lesssim \tilde{E} \lesssim 20$ TeV have been subjected to reduced adiabatic losses, compared to the particles at $\tilde{E} \lesssim 4$ TeV, leading to the more gradual transition at $r=0.9$.

The importance of diffusion relative to convection can be estimated using the \emph{P\'{e}clet} number \citep[e.g.][]{Prandtl1953}
\begin{equation}\label{eq:peclet} 
\xi = \frac{Vr}{\kappa}.
\end{equation}   
When $\xi \gg 1$ the system is convection dominated, while being diffusion dominated for $\xi \ll 1$.  Note that $\xi$ is in essence the ratio of the convection and diffusion time-scales, (\ref{eq:con_time}) and (\ref{eq:dif_time}).  With the ideal MHD limit imposed and assuming that $\kappa_r \propto 1/B$ (see Section \ref{sec:parameter}), the scaling $V = V_0(r_0/r)^{\beta}$ implies the scaling $\kappa = \kappa_0 (r/r_0)^{(1-\beta)}\kappa_p$.  Inserting these radial profiles into (\ref{eq:peclet}) leads to 
\bb\label{eq:peclet_2}
\xi = \frac{V_0r_0}{\kappa_0 \kappa_p},
\ee  
where it will be recalled from Section \ref{sec:parameter} that $\kappa_p=\tilde{p}/\tilde{p}_S$.  It thus follows from (\ref{eq:peclet_2}) that the value of $\xi$ is not uniquely defined for the system.  For the low energy part of the spectrum in Figure \ref{fig:conv-diff}, one has $\xi > 1$, while $\xi < 1$ at high energies.  The energy, $\tilde{E}_{\rm{dif}}=\tilde{p}_{\rm{dif}}c$, where the two transported process are equal can be found by setting $\xi=1$.  Using the values of $V_0$ and $\kappa_0$ chosen for the numerical calculations lead to the estimate $\tilde{E}_{\rm{dif}}\sim 75$ TeV.  This is in fair agreement with the values of $\tilde{E}_{\rm{dif}}\sim 100$ TeV at $r=0.1$ and $\tilde{E}_{\rm{dif}}\sim 40$ TeV at $r=0.9$, which once again confirms the validity of the numerical scheme.

\subsection{Convection-diffusion-synchrotron}

\begin{figure*}[!t]
   \centering
  \includegraphics[height=10cm]{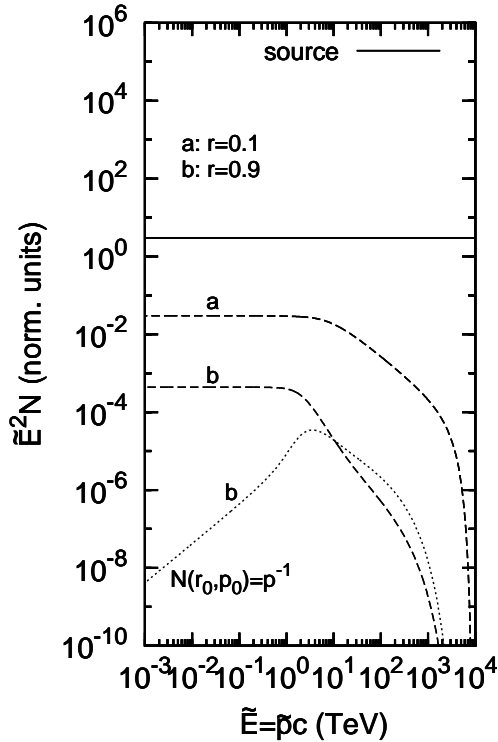}
   \caption{Numerical solutions for the convection-diffusion-synchrotron scenario.  Also shown is the spectrum at $r=0.9$ for the source spectrum $N(r_0,p_0)\propto p^{-1}$ (dotted line).}
              \label{fig:conv-diff-sync}
    \end{figure*}

The last step is to determine the effect of diffusion on the evolution of the spectra when synchrotron losses are present.  This section considers steady-state solutions to (\ref{eq:te}), with the transport processes of convection and diffusion included, together with adiabatic and synchrotron energy losses.  The results are plotted in Figure \ref{fig:conv-diff-sync}.

Compared to the convection system in Figure \ref{fig:conv-sync}, diffusion decreases the propagation time of particles through the system, thereby reducing the amount of synchrotron losses suffered by the particles.  The particles that lose the most energy due to synchrotron radiation are generally the particles with the largest diffusion coefficient, leading to a significant moderation of the characteristic synchrotron cut-off, as shown in Figure \ref{fig:conv-diff-sync}.

The spectrum at $r=0.1$ transitions from the adiabatic-dominated part at an energy $\tilde{E}\sim 5$ TeV, almost identical to the cut-off energy $\tilde{E}_{\rm{syn}} \sim 6$ TeV in Figure \ref{fig:conv-sync}.  The spectrum then softens to $N \propto \tilde{E}^{-3}$ at a slightly larger energy $\tilde{E} \sim 20$ TeV.  This is also the energy where the effect of diffusion becomes visible in Figure \ref{fig:conv-diff}.  A comparison of the diffusion-dominated part of the spectra shows that the intensity in Figure \ref{fig:conv-diff-sync} is lower than the corresponding intensity in Figure \ref{fig:conv-diff}.  The particles at $r=0.1$ in Figure \ref{fig:conv-diff-sync} have undergone a significant amount of synchrotron losses, but the shape of the spectrum is nevertheless determined by diffusion.  The exception is at very high energies, where a synchrotron cut-off appears.

At $r=0.9$ a more pronounced effect between the importance of diffusion, relative to synchrotron losses, can be seen.  The transition from the adiabatic-dominated part again occurs at an energy $\tilde{E} \sim 1$ TeV that is similar to $\tilde{E}_{\rm{syn}}\sim 1.4$ TeV in Figure \ref{fig:conv-sync}.  In the energy range $1 \mbox{ TeV}\lesssim \tilde{E} \lesssim 10$ TeV the spectrum initially softens,  and can be fitted with a power-law $N\propto \tilde{E}^{-3.8}$.  This initial softening of the spectrum is followed by a marginal harder spectrum at $\tilde{E} \sim 10$ TeV.  This is also the energy where the effect of diffusion becomes visible in Figure \ref{fig:conv-diff}.  In the energy range $10 \mbox{ TeV}\lesssim \tilde{E} \lesssim 200$ TeV the spectrum can be fitted with the power law $N \propto \tilde{E}^{-3.5}$.  At $\tilde{E} \gtrsim 100$ TeV the spectrum softens again,  signifying the beginning of the high-energy cut-off.  Although one might be inclined to believe that the effect of diffusion on synchrotron losses would only be important for $\tilde{E} \gtrsim 10$ TeV, Figure \ref{fig:conv-diff-sync} shows that diffusion already reduces the amount of synchrotron losses for $\tilde{E} \lesssim 10$ TeV.

Shown for comparison is the spectrum at $r=0.9$ for a $N\propto p^{-1}$ source spectrum.  For the purpose of clarity, this spectrum has been divided by a factor $10^3$.  The spectrum can be fitted with a power-law $N \propto \tilde{E}^{-3}$ in the energy range $10 \mbox{ TeV} \lesssim \tilde{E} \lesssim 200$ TeV.  

While the spectra shown if Figures \ref{fig:conv-sync}-\ref{fig:conv-diff-sync} were calculated using a free-escape outer boundary condition, it may be argued that a realistic scenario should have a partial escape boundary.  To investigate the influence of the outer boundary on the spectra, a no-escape condition ($\mathbf{S}=0$) was also imposed on the convection-diffusion-synchrotron scenario, where it was found that the choice of outer boundary condition does not affect the solutions.  When energy losses are present, particles not only escape in radial, but also momentum space, and the choice of outer boundary condition becomes unimportant.  This is especially true for the high-energy particles for which synchrotron losses are important.

Observations of the Vela PWN, located at a distance of $290$ pc \citep{Dodson2003}, show a bright X-ray nebula surrounding the termination shock \citep{Mangano2005}.  The synchrotron photon index, $\Gamma$, in the $3-10$ keV range was extracted from a number of annular regions of increasing size, revealing a softening of the index with increasing distance from the shock, located at $r_{\rm{ts}}=0.35'$ \citep{Ng2004}.  In the inner region, $r \le 0.5'$, the photon index was found to be $\Gamma=1.5\pm 0.02$, while in the outer annular region $8'\le r\le 12'$, the index was found to be $\Gamma=1.9\pm 0.06$.  If the energy spectrum of the particles is described by $N \propto E^{-\alpha}$, then the relation $\Gamma=(\alpha+1)/2$ implies $\alpha=2\pm 0.04$ in the inner region, and $\alpha=2.8\pm 0.12$ in the outer annular region.  A similar X-ray observation ($0.5-9$ keV) has also been performed for the nebula MHS 15-52, located at a distance of $5.2 \pm 1.4$ kpc \citep{Gaensler1999}.  In the inner annular region, $30''\le r \le 57''$, the photon index was was found to be $\Gamma=1.66\pm 0.02 \hspace{0.2cm}(\alpha=2.32\pm 0.04)$, softening to $\Gamma=2.24\pm 0.28 \hspace{0.2cm}(\alpha=3.48\pm 0.56)$ in the outer annular region, $246'' \le r \le 300''$ \citep{Schock2010}.       

The electron energy, $\tilde{E}$, needed to produce a synchrotron photon with a keV energy, $\tilde{E}_{\rm{keV}}$, is \citep{Dejager2009}
\bb\label{eq:min_energy}
\tilde{E} \approx (220\mbox{ TeV})\tilde{B}_{\mu\rm{G}}^{-1/2}\tilde{E}_{\rm{keV}}^{1/2}.
\ee
With the parameters chosen in Section \ref{sec:parameter}, the magnetic field has the values $\tilde{B}_{r=0.1}=113$ $\mu$G, and $\tilde{B}_{r=0.9}=38$ $\mu$G.  Inserting these values into (\ref{eq:min_energy}), one finds that the electron energy needed to produce synchrotron emission in the range $\tilde{E}_{\rm{keV}}=0.5-10$ keV is $\tilde{E}_{r=0.1}=15-65$ TeV and $\tilde{E}_{r=0.9}=25-113$ TeV.  Figure \ref{fig:conv-diff-sync} shows that these are the particle energy ranges where diffusion will have the largest influence on the evolution of the spectra.  Furthermore, the index $\alpha_{r=0.9}=3.5$ derived from Figure \ref{fig:conv-diff-sync} (in the energy range $2 \mbox{ TeV} \lesssim \tilde{E} \lesssim 100$ TeV) is consistent with the observed index in the outer regions of MSH 15-52.  Selecting a wind profile $V \propto 1/r^{0.1}$, while keeping all the other parameters fixed (see Section \ref{sec:parameter}), the model calculates a particle index $\alpha_{r=0.9}=2.8$, in agreement with the observed index in the outer regions of the compact Vela PWN.  

It should be kept in mind that the observations represent an average of the particle spectra in an annular region, whereas the results in Figure \ref{fig:conv-diff-sync} show spectra at a given point.  Even with this caveat in mind, the results from Figure \ref{fig:conv-diff-sync} indicate that diffusion should play an important role in determining the photon index, and can help to explain the observed softening of the spectrum.

\section{Time-dependent solutions}

Pulsar wind nebulae are dynamical systems, and the solutions presented thus far will only be applicable in the inner regions where the nebula has reached a (quasi) steady-state.  The time needed for the whole nebula to reach a steady-state may also be the time in which a realistic PWN has significantly expanded.  To take this expansion into account, it becomes necessary to solve (\ref{eq:te}) time-dependently.

With the added dimension of time, (\ref{eq:te}) is solved using the \emph{Douglas} Alternating Direction Implicit numerical scheme \citep{Douglas1962}, with time chosen as the stepping parameter.  This requires boundary conditions for both the radial and momentum direction.  For the radial direction, the boundary conditions used for the steady-state scheme are again imposed.  Since the present model only includes energy losses, there is a flow in momentum space from higher to lower momenta, similar to convection in configuration space.  Particles should be able to escape from momentum space when they reach $p_{\min}$, and the free-escape condition, $f(r,p_{\min})=0$, is imposed.  For $p_{max}$ the choice is not so clear, and the free-escape boundary is also imposed.  The grid-spacing is chosen as $\Delta r=2\times 10^{-3}$, $\Delta \ln p = 10^{-3}$, and for the stepping parameter, $\Delta t = 0.1$.  In order to make the time-dependent results directly comparable to the steady-state solutions, the same values for $V_0$, $B_0$, and $\kappa_0$ are used at the inner boundary of the system, unless explicitly stated otherwise.  Additionally, the radial profiles for $V$, $B$, and $\kappa$ are the same as in Section \ref{sec:parameter}.

In the first evolutionary phase \citep[e.g.][and references therein]{Gaensler2006}, the outer boundary of the nebula $\tilde{R}_{\rm{pwn}}$ expands with the convection velocity $\tilde{V}_{\rm{pwn}}$.  Theoretical calculations predict that $\tilde{R}_{\rm{pwn}} \propto t^{1.2}$ \citep[e.g.][]{Reynolds1984}, implying an expansion that accelerates with time.  A similar expansion rate has also been derived by e.g. \cite{Gelfand2009}, who find $\tilde{R}_{\rm{pwn}} \propto t^{1.1}$.  Furthermore, \cite{Gelfand2009} calculated an expansion velocity that increases from $\tilde{V}_{\rm{pwn}}=900$ km/s to $\tilde{V}_{\rm{pwn}}=2500$ km/s over a 4000 year timespan.  

Theoretical calculations also predict a time-dependence of the average magnetic field, ranging from $\bar{B} \propto t^{-1.3}$ to $\bar{B} \propto t^{-1.7}$ \citep[e.g.][]{Reynolds1984, Gelfand2009}, where this time-dependence is a direct result of the expansion of the system.  In one of the scenarios presented below, solutions are calculated for a static system with an explicit time-dependence in the magnetic field and diffusion coefficient.  The solutions presented for this system are not meant to represent the effect of an expanding boundary, but serve as an independent illustration of the effect of time-varying parameters.  The time-dependence $B \propto t_0/t$ is chosen for the magnetic field.  The scaling $\kappa \propto 1/B$ thus implies the time-dependence $\kappa \propto t/t_0$.

One possible way to take into account the effect of moving boundaries is to transform (\ref{eq:te}) into a coordinate system where the boundaries remain stationary.  An example of a general transformation is
\begin{equation}\label{eq:gen_trans}
r = \rho (r',t)r' + \epsilon (r',t),
\end{equation}
where $r$ is the coordinate in the expanding system, and $r'$ the coordinate in the transformed, static system.  For the present simulations, the expansion of the outer boundary is described by the transformation
\bb\label{eq:PWN_trans}
r = \left[V_{\rm{pwn}}\left(\frac{t-t_0}{r_1'-r_0'}\right)+1\right]r' + V_{\rm{pwn}}\left(\frac{t-t_0}{r_1'-r_0'}\right)r_0',
\ee
where $r_0'$ and $r_1'$ are respectively the inner and outer boundaries of the static system, and $t_0$ the time when expansion starts.  Comparison with (\ref{eq:gen_trans}) shows that the coefficient of $r'$ in (\ref{eq:PWN_trans}) can be identified with the variable $\rho$, and the second term on the right-hand side of (\ref{eq:PWN_trans}) with $\epsilon$.   

Using the transformations
\[
\frac{\partial f}{\partial t} =  \frac{\partial f'}{\partial t}-\frac{\partial r'}{\partial t}\frac{\partial f'}{\partial r'} = \frac{\partial f'}{\partial t}-\frac{1}{\rho}\frac{\partial r}{\partial t}\frac{\partial f'}{\partial r'}
\]
\begin{equation}
\frac{\partial f}{\partial r} = \frac{\partial r'}{\partial r}\frac{\partial f'}{\partial r'} = \frac{1}{\rho}\frac{\partial f'}{\partial r'},
\end{equation}
together with the fact that the distribution function is invariant 
\begin{equation}
f\left(r,t\right) = f\left(\rho r'+\epsilon,t\right) = f'(r',t),
\end{equation} 
transforms (\ref{eq:te}) to
\begin{multline}\label{eq:tte}
\frac{\partial f'}{\partial t} 
= \frac{\kappa}{\rho^2}\frac{\partial^2 f'}{\partial r'^2} + \frac{1}{\rho}\left[\frac{2\kappa}{\rho r'+\epsilon}+\frac{1}{\rho}\frac{\partial \kappa}{\partial r'}-V+\frac{\partial r}{\partial t}\right]\frac{\partial f'}{\partial r'}\\
 + \left[\frac{2V}{3(\rho r'+\epsilon)}+\frac{1}{3\rho}\frac{\partial V}{\partial r'}+z p\right]\frac{\partial f'}{\partial \ln p}+4zpf' + Q.
\end{multline} 
The derivative
\bb\label{eq:relate}
\frac{\partial r}{\partial t} = V_{\rm{pwn}}\left(\frac{r'-r'_0}{r_1'-r'_0}\right)
\ee
that appears in the coefficient of the term $\partial f/\partial r'$ in (\ref{eq:tte}) relates the velocity of the coordinate $r'$ to the expansion velocity $V_{\rm{pwn}}$.  At $r'=r_0'$ the derivative disappears, while being equal to $V_{\rm{pwn}}$ at $r'=r_1'$.  Note that at $r'=r_1'$, the derivative (\ref{eq:relate}) cancels with the convection velocity $V=V_{\rm{pwn}}$ in the coefficient of $\partial f/\partial r'$.

\subsection{Initially empty system}

To demonstrate the solutions obtained with the numerical scheme, the initial condition is chosen as an empty system.  For this simulation, only convection, diffusion, and adiabatic losses are taken into account.  Additionally, any time-dependence in the coefficients is neglected, while the outer boundary remains static.  The source is switched on at $t=0$, and the system is allowed to reach a steady-state.

\begin{figure*}[!t]
   \centering
  \includegraphics[height=10cm]{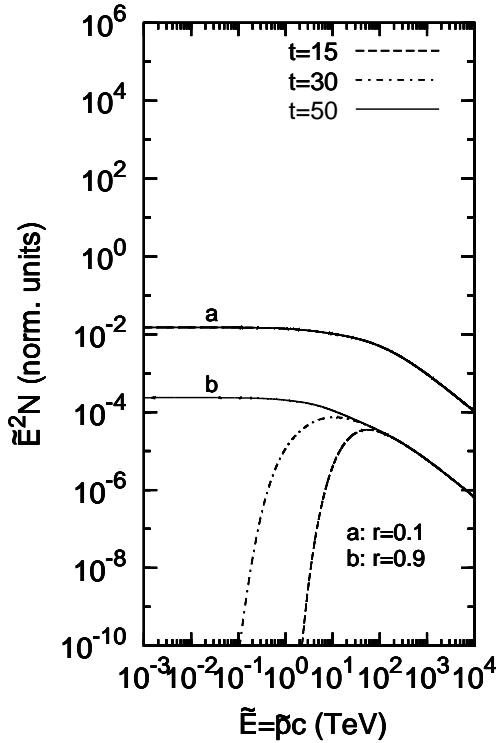}
   \caption{Time-dependent numerical solutions for an initially empty convection-diffusion system.  The system has reached a steady-state at $t=50$.}
              \label{fig:initial_empty}
    \end{figure*}

Figure \ref{fig:initial_empty} shows the evolution of the spectra at three different times, with the system having reached a steady-state at $t=50$.  At times $t=15$ and $t=30$, the spectra at $r=0.1$ already resemble the steady-state spectrum (cf. Figure \ref{fig:conv-diff}).  This is also true for the high-energy part of the spectra at $r=0.9$.  However, at lower energies the spectra develop a cut-off that decreases in energy with increasing time.  Particles with energies below this cut-off cannot propagate as effectively to the outer parts of the system as a result of a small diffusion coefficient.  The energy where the cut-off should appear can be estimated using the diffusion time-scale, (\ref{eq:dif_time}).  Using the average value of $\kappa$ between $r_0$ and $r=0.9$ leads to the estimate $\tilde{E}\sim 40$ TeV for $t=15$, and $\tilde{E} \sim 20$ TeV for $t=30$, in good agreement with the values shown in Figure \ref{fig:initial_empty}.  This demonstrates the soundness of the numerical scheme.

\subsection{Time-dependent coefficients}

In this section the effect of time-dependent coefficients on the evolution of the spectra is illustrated.  The processes of convection, diffusion, as well as both energy loss mechanisms, i.e. adiabatic expansion and synchrotron radiation, are included.  An initially empty system is allowed to reach a steady-state at $t_0=50$ (cf. Figure \ref{fig:conv-diff-sync}), before the time-dependence $B_0 \propto 1/t$ and $\kappa_0 \propto t$ is introduced.

\begin{figure*}[!t]
   \centering
  \includegraphics[height=10cm]{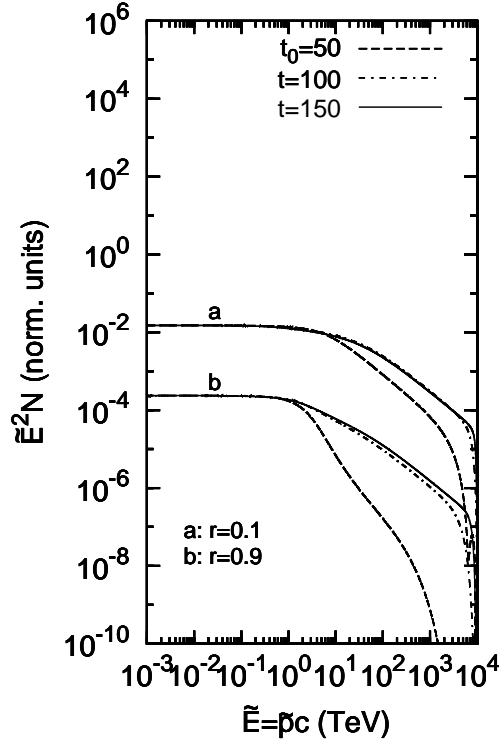}
   \caption{The effect of a time-dependent magnetic field that scales as $B\propto 1/t$, and a diffusion coefficient that scales as $\kappa \propto t$.  The time-dependence is implemented at times $t>t_0$, where $t_0=50$ represents the steady-state solution.  The solutions shown are for a convection-diffusion-synchrotron system.}
              \label{fig:B_k_time_dependent}
    \end{figure*}

Figure \ref{fig:B_k_time_dependent} shows the spectra as a function of time, for $t=100$ $(2\times t_0)$ and $t=150$ $(3\times t_0)$.  At $r=0.9$ the time-dependence is much more pronounced than at $r=0.1$.  With $B \propto 1/t$, the synchrotron loss rate, (\ref{eq:a_coef}), decreases as $z \propto 1/t^2$, while the escape time, (\ref{eq:dif_time}), decreases as $\tau_{\rm{dif}}\propto 1/t$.  Since the amount of synchrotron losses is dependent on both $z$ and $\tau_{\rm{dif}}$, the change in the amount of synchrotron losses suffered is $\propto 1/t^3$.  From $t_0=50$ to $t=100$ there is a rapid increase in the intensity, but from $t=100$ to $t=150$ the synchrotron losses have become so small that the intensity saturates.

\subsection{System with expanding boundary}

\begin{figure*}[!t]
   \centering
  \includegraphics[height=10cm]{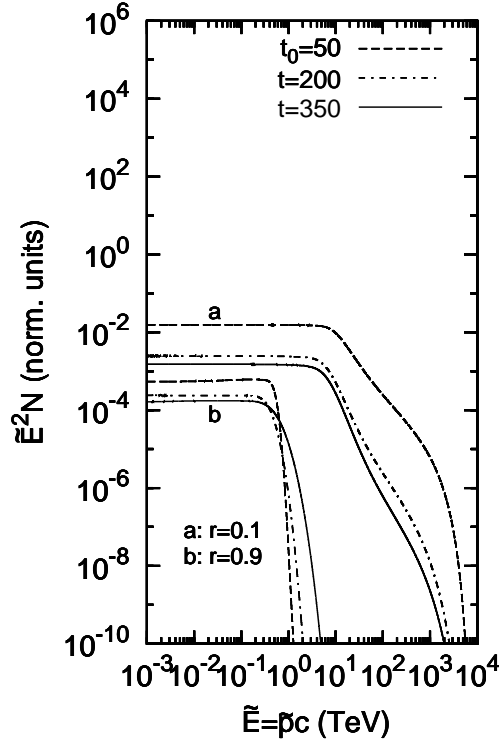}
   \caption{Time-dependent solutions for a convection-diffusion-synchrotron system that expands at $t>t_0$.}
              \label{fig:moving_bound}
    \end{figure*}

The last scenario demonstrates the effect of a moving boundary.  Figure \ref{fig:moving_bound} shows the solutions of (\ref{eq:tte}) with the processes of convection, diffusion, adiabatic expansion, and synchrotron losses included.  The velocity at the inner boundary is $\tilde{V}_0=0.3c$, while a constant expansion velocity, $\tilde{V}_{\rm{pwn}}=2000$ km/s, is chosen.  The inner and outer boundaries are at $r_0=0.01$ and $r=1$, respectively, with the initial size of the system chosen as $\tilde{r}_S=1$ pc.  The velocities at the boundaries imply a radial convection profile $V \propto 1/r^{0.83}$.  

The value $\tilde{B}_0=500$ $\mu$G is chosen at the inner boundary of the system, with the radial scaling of the magnetic field determined using the ideal MHD limit, (\ref{eq:MHD_limit_reduce}).  The radial profile of the magnetic field is therefore given by $B \propto 1/r^{0.17}$.  The diffusion coefficient has a value $\kappa_0=3 \times 10^{-5}$, or in terms of the unscaled value, $\tilde{\kappa}_0=10^{24}$ $\rm{cm}^2/\rm{s}$.  This is very similar to the value $\kappa_0=4 \times 10^{-5}$ chosen for the steady-state simulations.  The radial scaling of $\kappa_r$ is once again chosen as inversely proportional to the scaling of the magnetic field, implying $\kappa_r \propto r^{0.17}$.

To find an appropriate initial condition, the system is first allowed to reach a steady-state at $t_0=50$.  Starting at $t=t_0$, the system then expands until $t=350$, representing a factor seven increase in the size of the system.  Note that the spectra shown at the different times are plotted at the positions in the transformed, static system, i.e. at $r'=0.1$ and $r'=0.9$.  

The first noticeable feature in Figure \ref{fig:moving_bound} is that the intensities in the convection-dominated part of the spectrum decrease with time.  This is a direct result of the expansion of the system.  Adiabatic losses scale as $V/r$ in a spherical system, leading to larger losses in the inner part of the system.  This is also the reason why the amount of adiabatic losses at a given position in the static system, $r'$, decreases with time.  As $t$ increases, $r'$ represents larger $r$ values in the PWN, leading to less adiabatic losses.   

In the diffusion/synchrotron dominated regime, the spectra at $r'=0.1$ become softer with increasing time, while the synchrotron cut-off at $\tilde{E}\sim 5000$ TeV shifts to lower energies.  For increasing values of $t$ the particles have travelled further through the system, and have therefore been subjected to more synchrotron losses.  In the first time interval, $50 \le t \le 200$, the amount of losses is still significant.  However, the decrease in $z$ is of such a nature that the amount of synchrotron losses is significantly reduced in the second time interval, $50 \le t \le 200$.  

At $r'=0.9$ the spectra become marginally harder with time.  Furthermore, the largest increase occurs in the second time interval.  To understand this opposite behaviour, it is important to take into account all the contributing factors.  Firstly, if the value of the diffusion coefficient remains unchanged, then a larger system leads to longer residence times, thereby reducing the relative importance of diffusion.  However, this effect is partially offset by the fact that the system expands at a constant velocity.  At times $t \ge t_0$ the velocity is described by the profile $V\propto 1/r^{\beta}$, with $\beta=0.83$.  Since the velocity at the inner and outer boundaries remains constant, the index of the radial profile has evolved to $\beta=0.58$ at $t=350$.  The scaling $B \propto 1/r^{1-\beta}$, together with the scaling $\kappa \propto 1/B$, implies that the diffusion coefficient becomes larger in the outer part of the system with increasing time, thereby counteracting the effect of longer residence times.  Although the change in the index of the convective radial profile implies an identical change in the index of the magnetic field profile, the amount of synchrotron losses as a function of position decreases more rapidly.  This is due to the $z\propto B^2$ dependence of the synchrotron loss coefficient, (\ref{eq:a_coef}), leading to a change in the profile of $\Delta \beta = 0.48$.  The effect of synchrotron losses therefore decreases faster with time, becoming less important in the outer parts of the system.  

In the absence of diffusion, one would expect expect the synchrotron cut-off to shift to lower energies with an increase in time.  However, since the importance of diffusion increases with time, the effect is a marginal hardening of the spectrum.  This hardening of the spectrum at $t=350$ is not caused by particles that were originally located at $r'=0.9$ at $t_0=50$, but rather by particles that have diffused from distances $r' < 0.9$ to $r'=0.9$ at times $t > 50$.  

The spectra at $t_0=50$ are the solutions that would have been found if (\ref{eq:te}) were solved time-independently, i.e. $\partial f/\partial t = 0$.  Thus, in summary, the moving boundary predominantly leads to a reduction in the intensity and radial particle gradients with time, but has a limited influence on the shape of the spectrum.

\section{Discussion and Conclusions}

This paper has demonstrated a simple and workable numerical solution to the Fokker-Planck transport equation in a spherically-symmetric, central source system.  The archetypical example of such a system is a pulsar wind nebula, but other possible examples include globular clusters and the non-thermal bubbles observed above and below the Galactic plane.  The main contribution of the paper is to show that the inclusion of diffusion transforms the well-known synchrotron cut-off in the spectrum to a much more gradual roll-over at high energies.  It has been observed from a number of PWN'e that the particle spectrum softens with increasing distance form the source \citep[e.g.][]{Mangano2005, Schock2010}.  Traditionally attributed to synchrotron losses alone, it is found that the softening can be better explained as a result of the aforementioned roll-over of the spectrum.  

An important process that was investigated in this paper is the effect that an expanding outer boundary has on the evolution of the system.  The primary effect of this dynamical process is to reduce the importance of diffusion.  This decrease can be related to the fact that particles will have a longer residence time in a larger system.  It is important to note that the moving boundary as such does not influence the value of the diffusion coefficient.  However, the expansion of the system leads to a decrease in the magnetic field, which in turn leads to an increase in the value of the diffusion coefficient.  This increase in the diffusion coefficient is similar to the effect caused by increased residence times, and diffusion remains important for spectral evolution.  It was further found that an expanding outer boundary has a limited effect on the shape of the spectra   

Results from MHD simulations \citep[e.g.][]{Komissarov2003,Delzanna2004} show that the dipole structure of the pulsar's magnetic field is preserved downstream of the termination shock, leading to the formation of a current sheet \citep[e.g.][and references therein]{Kirk2009} similar to the one observed in the heliosphere \citep[e.g.][]{Smith2001}.  Observations and simulations indicate that this dipole structure of the magnetic field, along with the presence of the current sheet, leads to large-scale particle drifts in the heliosphere \citep[e.g.][and references therein]{Potgieter1985}.  It is natural that a similar effect should be present in pulsar wind nebulae, and a more realistic model should take this into account.  These drift effects can be included by expanding the present model to a higher-dimensional model that calculates the solutions not only as a function of radial distance, but also as a function of polar angle.  Additionally, the two-dimensional model has the advantage that more complicated flow patterns can be included.  A future paper is planned where these, multi-dimensional effects on the evolution of the particle spectrum will be demonstrated.


\bibliographystyle{aa}
\bibliography{references}

\end{document}